\documentclass{article}
\usepackage{spconf,amsmath,graphicx}
\usepackage{amssymb}
\usepackage[table]{xcolor}
\usepackage{multirow}
\usepackage{multicol}

\title{Progressive Unsupervised Domain Adaptation for ASR using Ensemble Models and Multi-stage Training}

\name{Rehan Ahmad, Muhammad Umar Farooq, Thomas Hain
\thanks{© 2024 IEEE.  Personal use of this material is permitted. Permission from IEEE must be obtained for all other uses, in any current or future media, including reprinting/republishing this material for advertising or promotional purposes, creating new collective works, for resale or redistribution to servers or lists, or reuse of any copyrighted component of this work in other works.}
}
\address{Speech and Hearing Group, The University of Sheffield, UK}

\begin{document}
\ninept
\maketitle
\begin{abstract}
In Automatic Speech Recognition (ASR), teacher-student (T/S) training has shown to perform well for domain adaptation with small amount of training data. However, adaption without ground-truth labels is still challenging. A previous study has shown the effectiveness of using ensemble teacher models in T/S training for unsupervised domain adaptation (UDA) but its performance still lags behind compared to the model trained on in-domain data. This paper proposes a method to yield better UDA by training multi-stage students with ensemble teacher models. Initially, multiple teacher models are trained on labelled data from read and meeting domains. These teachers are used to train a student model on unlabelled out-of-domain telephone speech data. To improve the adaptation, subsequent student models are trained sequentially considering previously trained model as their teacher. Experiments are conducted with three teachers trained on AMI, WSJ and LibriSpeech and three stages of students on SwitchBoard data. Results shown on eval00 test set show significant WER improvement with multi-stage training with an absolute gain of 9.8\%, 7.7\% and 3.3\% at each stage.
\end{abstract}
\section{Introduction}
\label{intro}
\vspace{-3mm}

The performance of automatic speech recognition (ASR) has been improved dramatically with advanced end-to-end neural network architectures \cite{li2022recent}. However, it is shown that these models does not perform well \cite{likhomanenko2020rethinking} on cross-domain data due to mismatch between training and test data domains. Domain adaptation \cite{bell2020adaptation} tries to solve this problem by transferring the model trained on source data to the target domain data. Adaptation is achieved via supervised \cite{sim2018domain} or unsupervised \cite{khurana2021unsupervised} methods depending on the availability of labelled target data. Domain adaptation method can be straightforward with the presence of labelled target data i.e. fine-tuning the source models on the target data \cite{sim2018domain, hou2021exploiting}. However, with unlabelled target data unsupervised domain adaptation (UDA) requires specific modelling \cite{manohar2018teacher, khurana2021unsupervised} approaches to efficiently learn the new domain. 

Teacher-student (T/S) \cite{KD} training is a common approach to transfer knowledge of one model to another. It has been widely used for model compression \cite{chebotar2016distilling, kim2019knowledge, takashima2018investigation}, domain adaptation \cite{bell2020adaptation,asami2017domain,meng2019domain,zhu2020domain, zhang2020semi} and generalisation \cite{kim2021domain}. 
Ensemble teacher models in T/S training have shown great benefit for acoustic model training \cite{fukuda2017efficient, gao2021distilling, li2017semi}. Having multiple teachers, a student can adapt better \cite{ahmad2023towards} leveraging diverse teacher outputs. The output of multiple teacher models can be effectively combined to train the student model. One of the simplest approach of combining teacher outputs is taking the average of the posterior distributions from all the teachers \cite{chebotar2016distilling, fukuda2017efficient}. However, previous work \cite{gao2021distilling,ahmad2023towards} has shown that taking the average of posterior distributions from all the teacher models and using them in training a student ASR model does not lead to any improvements in error rate. 
In \cite{gao2021distilling}, authors have tested many selections methods including Top-1, Top-K, averaging and weighted-average with Top-1 being their best approach. Similarly, in \cite{ahmad2023towards} we have proposed Top-1 based unsupervised selection method for UDA to select the best posteriors from ensemble of teachers. 
With the proposed selection method student model is only trained on posteriors with lower error. Hence, the student model performed significantly better compared to all the teachers. However, the performance of the student model still lags behind compared to the state-of-the-art in-domain model.

Recently, iterative \cite{khurana2021unsupervised, xu2020iterative} and multi-stage \cite{rathod2022multi} training methods have been proposed for acoustic model training. In \cite{rathod2022multi}, Rathod et al. has showed that the multi-stage training can be used in T/S method to gradually compress the model without losing significant performance. Similarly, iterative pseudo-labelling \cite{xu2020iterative} and self-iteration methods \cite{khurana2021unsupervised} have shown improvements in semi-supervised and unsupervised data selection for ASR. Inspired by improvements shown in these methods, multi-stage training is exploited in this work to progressively improve domain adaptation. 

This paper improves previously proposed UDA in ensemble T/S training method for ASR \cite{ahmad2023towards} by integrating multi-stage training. In our previous work \cite{ahmad2023towards}, multiple teacher models were trained on the source data. These teachers were used to generate soft-labels (posteriors) for the target data during inference. From multiple teachers, soft-labels were selected by unsupervised selection method to train a student model. The results showed significant WER improvement on target data by the student model compared to the teachers. To improve the adaptation, several changes are made in this work which is explained as follows. 
Since beam-search \cite{ABDOU2004409} has been proven to perform better than greedy decoding, therefore in this work soft-labels generate by teacher models are replaced by hard-labels by decoding through beam-search with out-of-domain (OOD) LM. The student model is now trained on the hard-labels.
Moreover, as discussed in the previous work that the student performed better than the teachers, therefore in this work the trained student is used as a teacher to improved previously generated pseudo-labels. The new pseudo-labels are then used to train another student model. The process of generating the pseudo-labels by previous student and using them to train next student is repeated to progressively improve the adaptation. The experiments are conducted with three source (labelled) data from meeting and read speech and a target (unlabelled) data from conversational telephone speech. The results show that the student models trained up to three stages show significant WER improvement at each stage with an absolute difference of 9.8\%, 7.7\% and 3.3\%. To study some insights, a comparative analysis of fine-tuning wav2vec-base and wav2vec-large models is conducted at each training stage. Secondly, WER of pseudo-labels generated by teacher and student models is also observed at each training stage.

\vspace{-2mm}
\section{Ensemble and Multi-stage models}
\label{sec:multiStage}
\vspace{-2mm}
The proposed ensemble teacher-student training paradigm assumes the existence of $N$ teacher models trained independently on different labelled datasets. Each model is based on a pre-trained wav2vec2.0 \cite{baevski2020wav2vec}, connected with feed forward network (FFN) layers. Input to the wav2vec2.0 models is a raw waveform represented by $X$ and outputs are the feature representations $\mathbf{c_1,...,c_T}$ for $T$ time-steps. These features are fed to the FFN layers to produce the output posterior distributions $\mathbf{h_1,...,h_T}$. Each $\mathbf{h_i} \in \mathbb{R}^z$ with a grapheme vocabulary represented by $G=\{g_1, g_2,...g_z\}$. All the teacher and student models are fine-tuned using CTC \cite{graves2012connectionist} loss defined as:

\begin{equation}
\label{eq1}
    \mathcal{L}_{CTC} = -\sum_B \log p(\mathbf{y}|\mathbf{h})
\end{equation}
where $\mathbf{y}$ represents the grapheme and $B$ is the whole training dataset. During fine-tuning, teacher models use ground-truth labels while students use pseudo-labels generated by the teacher(s).

Outputs of the ensemble teachers are used to train a single student model which is further extended into multi-stage models. In multi-stage training, $M$ student models are trained successively using pseudo-labels generated by the last stage model. The value of $M$ depends on the progressive improvement at each stage which is kept increasing until no improvement is observed in the WER of a subsequent student. Each stage of the student is represented by $Sk$, where $k \in \{1,2,...,M\}$. During inference, each teacher model generates output posterior distributions for the same utterance, they are selected via unsupervised sampling method to train the student. The following section describes the selection strategy in detail. Figure \ref{fig:model} shows the modelling approach comprising of ensemble teacher models and multi-stage students.

\vspace{-2mm}
\subsection{Selection strategy}
\label{subsec:selection}
An unsupervised selection strategy is employed for the ensemble teachers to select the best teacher's output. The selection is made on the basis of average posterior values from each teacher model. In \cite{ahmad2023towards}, we discussed many selection strategies to select the best output from ensemble of teachers including `averaging', `frame-wise' and `Top-1'. However, the Top-1 strategy has been proved to outperform all other techniques and is employed in this work. 

For each input utterance, $N$ teacher models generate output posterior distributions for $T$ time-steps represented as $\mathbf{h_{1}^1...h_{T}^N}$. At each time-step $i$ maximum posterior value for teacher $n$ is selected as:

\begin{equation}
    p_{i}^n = \max_z \mathbf{h_{i}^n}
\end{equation}
where $z$ is the vocabulary size.

An average is taken across the time-steps to get an averaged posterior value for each teacher as:

\begin{equation}
\label{eq:avgpost}
    q^n = \frac{1}{T} \sum_{i=1}^{T} p_{i}^n
\end{equation}
This average value $q^n$ is the score of each teacher $n$ and the teacher with maximum score is selected as: 

\begin{equation}
\label{eq4}
    b = \arg \max_n q^n
\end{equation}
where $b$ is the selected teacher number. The output posterior distributions of teacher $b$ is used to train the student model. With this selection strategy output posterior distribution for each utterance is dynamically selected from one of the teacher models on the basis of their scores.

\vspace{-2mm}
\subsection{Training method}
\begin{figure}[t]
  \centering
  \includegraphics[width=\linewidth]{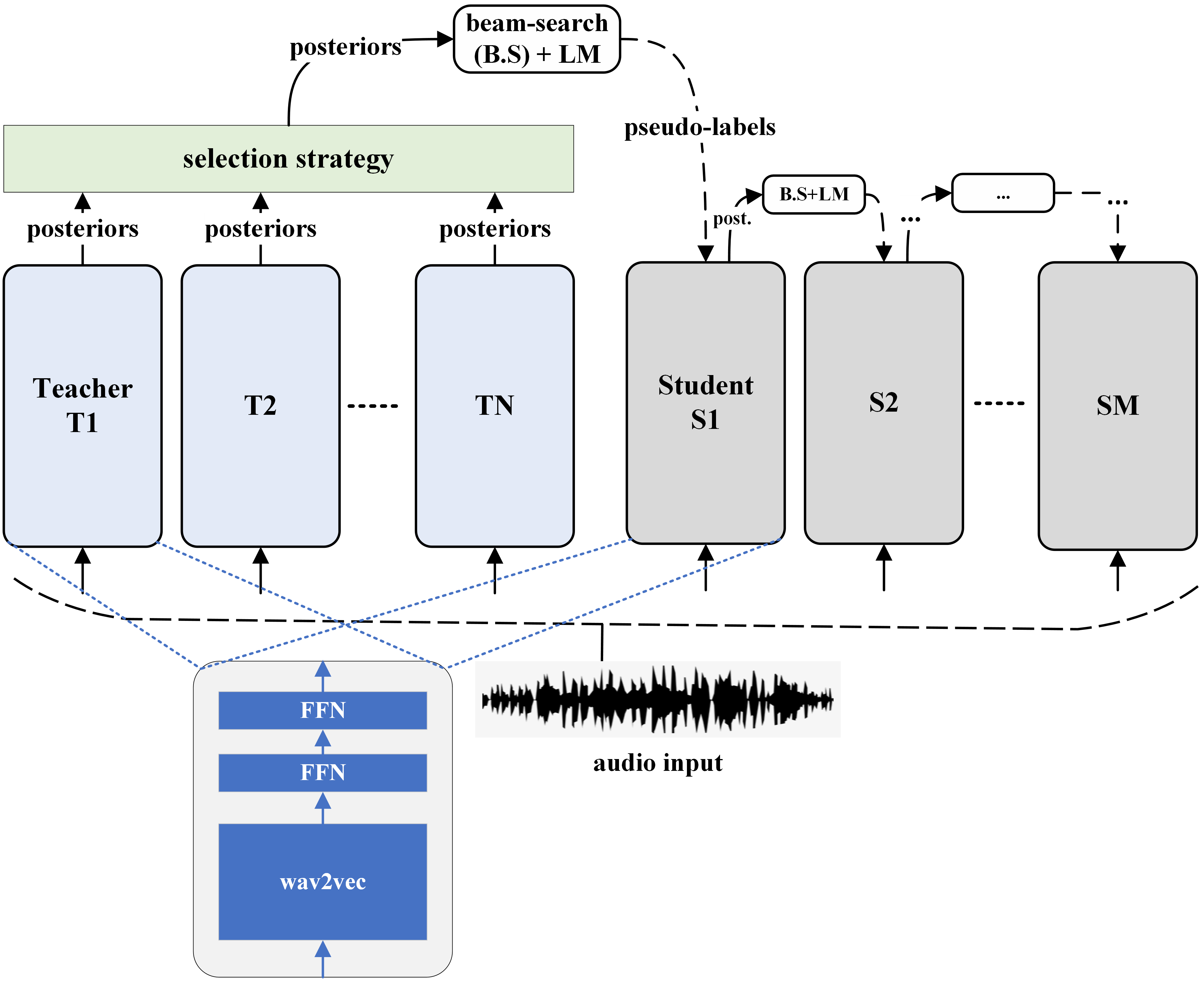}
  \caption{An ensemble of teacher models with multi-stage students. Each teacher and student model is based on wav2vec with Feed Forward Network (FFN) layers. Each model produces posterior distributions during inference and takes the pseudo-labels during fine-tuning.}
  \label{fig:model}
  \vspace{-3mm}
\end{figure}

Figure \ref{fig:model} shows the ensemble teacher models with multi-stage students. $N$ teacher models are trained independently with $N$ source domain labelled datasets. 
To train a student model, an unlabelled out-of-domain (OOD) target data is used. Since the target data is unlabelled, therefore student is trained using the labels generate by the teacher models. 


To train the $S1$ student, all the teachers run inference on the target data and generate output posterior distributions. The posteriors of the best teacher are selected by the selection strategy discussed in the Section \ref{subsec:selection}. Selected posteriors are finally converted into pseudo-labels using beam-search\footnote{https://github.com/kensho-technologies/pyctcdecode} integrated with an out-of-domain LM as:
\vspace{-1mm}
\begin{equation}
\label{eq:beamsearch}
    \hat{\mathbf{y}} = \arg \max_\mathbf{y}  \log p_{AM}(\mathbf{y}|\mathbf{x}) + \alpha \log p_{LM}(\mathbf{y}) + \beta |\mathbf{y}|
\end{equation}
where $p_{AM}$ and $p_{LM}$ represent acoustic and language model probabilities. $\alpha$ and $\beta$ are hyper-parameters optimised on validation set. 

In training the $S1$ model, the supervision is provided by ensemble teacher models. As the $S1$ model training benefited from diverse teacher outputs, therefore it outperforms all the teacher models. This has already been shown in \cite{ahmad2023towards}.
Once $S1$ is trained, it is used as a teacher to update previously generated pseudo-labels. 
The updated pseudo-labels are then used to train next stage student $S2$. This process is repeated for $M$ number of stages and training process is stopped when no improvement is noticed in a subsequent student model. 
The idea behind multi-stage training is that if a student performs better than its teacher then use this student as a teacher to update the previous pseudo-labels. Then, train a next student with the updated labels. 

\section{Experiments}
\vspace{-2mm}
\subsection{Data}
\vspace{-2mm}
Four corpora from different domains have been used to experiment with ensemble multi-stage teacher-student training. These include AMI \cite{Carletta2006}, WSJ \cite{paul1992design} (LDC93S6A, LDC94S13A), LibriSpeech (LS) \cite{Panayotov2015} and SwitchBoard (SWBD) \cite{Godfrey1992}. From LibriSpeech, a smaller subset of 360h is taken which is referred as LS360. 
Durations of AMI, WSJ and SWBD is 100, 272 and 300 hours respectively. SwitchBoard audio data is upsampled to 16kHz for wav2vec2.0 model requirement. In terms of domains, AMI is meeting speech recorded in a meeting room, WSJ and LS360 are read speech data recorded in the studio and SWBD is conversational telephone speech. Among these corpora AMI, WSJ and LS360 are used as source labelled data to train teacher models and SWBD as an unlabelled taget data for the student. Student models are evaluated on the eval00 test set. This test set combines CallHome (LDC97S42) and SwitchBoard test sets represented as CH and SB.
\vspace{-2mm}

\subsection{Experimental setup}
In the experiments, three teacher models are considered, i.e $N=3$, and the student models are trained up to three stages $M=3$. Teacher models are based on wav2vec-large pre-trained on LV-60K and fine-tuned independently on LS360, WSJ and AMI. For the student model, either wav2vec-base pre-trained with LS960 or wav2vec-large pre-trained with LV-60K is considered for different set of experiments. The student is fine-tuned on SWDB pseudo-labels whose domain is completely mismatched with the teachers. To acquire pseudo-labels and evaluate each student model, a 3-gram language model is trained by combining the transcripts of AMI, WSJ and LS360. As LM is trained on the teachers data, it is also an out-of-domain model for the student.  

The first experiment investigates the advantage of training the student model using hard-labels generated by beam-search with OOD LM. In this experiment, ensemble teacher models are used to train the $S1$ student model. 
The WER of $S1$ model is compared with $S_{KL}$ and $S_{Or}$. $S_{KL}$ is the first-stage student trained on the soft pseudo-labels using KL loss. $S_{Or}$ is the first-stage student model trained with the oracle selection of the teachers. Oracle selection is made by selecting each posterior distribution which has lowest WER on each training utterance.  


In the second experiment, multi-stage training is conducted where $S1$ model is used as a teacher to generate the pseudo-labels of the training data and update the previously generated pseudo-labels. The $S2$ student is then trained using new labels. Similarly, subsequent $S3$ student is trained after $S2$ in a similar manner. All the models are evaluated on eval00, CH and SB test sets with and without using 3-gram language model. In the above experiments, all the models are based on wav2vec-large for the sake of fair comparison. Table \ref{tab:multi-stage} shows the WER of all the teachers and students. 

The third experiment provides a comparison between wav2vec-base and wav2vec-large self-supervised models which are fine-tuned on the noisy pseudo-labels at each stage of the student model. This experiment aims to analyse if large models with higher capacity in the number of parameters and pre-training data provide any gain compared to the smaller ones when trained with noisy labels. 
The wav2vec-base model is pre-trained with LS960 and has 95M parameters and wav2vec-large is pre-trained with LV-60K and has 317M parameters. These results are presented in Table \ref{tab:w2vcomp}. 

\begin{table*}[ht]
    \centering
    \footnotesize
    \caption{WER(\%) evaluation of teachers and multi-stage students on SWBD test sets with and without (OOD) LM. The shaded boxes show in-domain results of SWBD and rest are out-of-domain. $S_{Or}$ is the first stage student trained with oracle selection of posteriors and $S_{KL}$ is the first stage student trained on soft-posteriors (acquired through selection strategy) with KL loss. All the models are based on wav2vec-large.}
    \label{tab:multi-stage}
    \begin{tabular}{l|c|c|c|c|c c c c c}
        \hline
        \hline
         & \multicolumn{3}{c|}{ \multirow{2}{*}{\bf Teacher models}} & \bf Baseline & \multicolumn{5}{c}{\multirow{2}{*}{\bf Student models}} \\
         & \multicolumn{3}{c|}{} & \bf (in-domain) & \multicolumn{5}{c}{}\\
         \hline
        \multirow{2}{*}{\bf Test sets} & \multirow{2}{*}{\bf AMI} & \multirow{2}{*}{\bf LS360} & \multirow{2}{*}{\bf WSJ} & \multirow{2}{*}{\bf SWBD} & \multicolumn{4}{c}{\bf SWBD}\\
         & & & & & $S_{Or}$ & $S_{KL}$ & $S1$ & $S2$ & $S3$ \\
        \hline
        & \multicolumn{9}{c}{\textit{w/o LM}}\\
        \hline
        eval00 & 47.4 & 41.8 & 64.2 & \cellcolor{lightgray} 11.9 & 27.4 & 36.2 & 32.0 & 24.3 & \bf 21.0 \\
        CH & 52.0 & 46.8 & 71.6 & \cellcolor{lightgray} 15.3 & 31.6 & 40.3 & 36.0 & 28.1 & \bf 24.5 \\
        SB & 42.5 & 36.7 & 56.5 & \cellcolor{lightgray} 8.4 & 23.1 & 31.8 & 27.8 & 20.3 & \bf 17.4 \\
        \hline
        & \multicolumn{8}{c}{\textit{w/ LM (OOD)}}\\
        \hline
        eval00 & 44.3 & 38.2 & 61.8 & \cellcolor{lightgray} 10.1 & 22.4 & 28.5 & 26.2 & 21.3 & \bf 19.6 \\
        CH & 49.0 & 43.2 & 69.7 & \cellcolor{lightgray} 12.8 & 26.1 & 33.3 & 30.2 & 25.0 & \bf 23.1 \\
        SB & 39.3 & 33.0 & 53.5 & \cellcolor{lightgray} 7.3 & 18.5 & 23.6 & 22.0 & 17.4 & \bf 16.0 \\
        \hline
        \hline
    \end{tabular}
    \vspace{-4mm}

\end{table*}

\section{Results and Discussions}
\vspace{-2mm}

Table \ref{tab:multi-stage} shows the results of the first two experiments. We first report the WER for the three teacher models which are represented by the name of the dataset they were trained on i.e. AMI, LS360 and WSJ. These teachers provide state-of-the art results on their in-domain test sets \cite{ahmad2023towards}.
SWBD baseline results are also presented to show the in-domain results on eval00. For SWBD baseline, a wav2vec model is fine-tuned using labelled SWBD training data. 
The in-domain results are shown in shaded cells. All the results in this table are presented with and without an OOD LM. Same LM is used to in beam-search to produce pseudo-labels for each student. Among three teacher models it can be seen that the lowest WER on eval00, CH, SB is provided by LS360 model i.e. 41.8\%, 48.6\% and 36.7\% respectively, which is further improved by OOD LM upto 3.7\% absolute. Overall, for eval00 the LS360 model outperform AMI and WSJ by 5.6\% and 22.4\% reduction in WER.   


As the $S1$ model is trained by selecting the best teacher's output using selection strategy, a significant improvement is noticed compared to all the teachers with WER of 32\%. For $S1$, the absolute WER improvement over the best teacher (LS360) is 9.8\%, 10.8\%, and 8.9\% for eval00, CH and SB evaluation sets.  
Compared to the $S_{KL}$ model, which uses soft posteriors and KL loss in training first stage student, the $S1$ model outperforms by 4.2\% WER for eval00. This shows the significance of first proposed improvement in the training method by using beam-search with LM to get better training labels for the student model. The selection strategy is an unsupervised method which means there is a chance to select wrong teacher for each utterance therefore an oracle student model $S_{Or}$ is trained to compare with $S1$. The WER of the $S_{Or}$ is 4.6\% better than the $S1$, which shows there is a room to improve the selection method however the main challenge is an unsupervised way of selecting best teacher for each utterance. OOD LM lead to further improvement in the WER of the $S1$ from 32.0\% to 26.2\% for eval00 i.e. 5.8\% absolute gain. Similar, results are shown by CH and SB evaluation sets.

The $S1$ model outperforming all the teachers is used to update the pseudo-labels. The $S2$ model is then trained with new pseudo-labels. From table \ref{tab:multi-stage}, it can be seen that the $S2$ outperforming the $S1$ model with WER of 24.3\% (w/o LM) and 21.3\% (w/ OOD LM) for eval00. The absolute gain by the $S2$ is 7.7\% (w/o LM) and 4.9\% (w/ OOD LM). Similarly, the $S3$ model trained on pseudo-labels provided by the $S2$ outperforming it with WER of 21\% (w/o LM) and 19.6\% (w/ OOD LM) with gain of 3.3\% (w/o LM) and 1.7\% (w/ OOD LM). Comparing the individual gain at each stage of the student models it is observed that the highest improvement is noticed at first stage and then it gradually decreases as the performance of the model gets closer to the in-domain baseline.
In brief, the WER improvement on eval00 at each stage compared to their teachers is 9.8\%, 7.7\% and 3.3\%. WER reduction from LS360 to $S3$ is 20.8\% and $S1$ to $S3$ is 11\%. These results show the significance of the proposed methods in improving the unsupervised domain adaptation. Compared to our previous method \cite{ahmad2023towards}, WER of the multi-stage are much closer to the SWBD baseline model.


Table \ref{tab:w2vcomp} presents the results of the third experiment where the wav2vec-base (w2v-B) and the wav2vec-large (w2v-L) models are trained at each stage and their WER are compared. 
For the $S1$ model in table \ref{tab:w2vcomp}, the results show that the w2v-L model do not show improvements on eval00 and SB test sets and minimal improvement on CH. On eval00, WER of the w2v-B is 0.5\% better than the w2v-L w/o LM and 0.2\% w/ LM. Similar difference is noticed for SB test set. However, for CH set the w2v-L is 0.4\% better than the w2v-B. At the first stage training of the student model the results show that there is no benefit of using the w2v-L model over the w2v-B model. 
For the $S2$, some improvements are noticed for the w2v-L compared to the w2v-B with 0.4\% (w/o LM) and 0.9\% (w/ OOD LM) gain in WER on eval00 and 1.1\% (w/o LM) and 1.7\% (w/ OOD LM) on CH. This difference is further improved by the $S3$ model with 1.9\%, 3.3\% on eval00 and CH.
The results show that as the subsequent model gets better compared to its teacher and provide better pseudo-labels for the next stage, the w2v-L model starts showing improvement over the w2v-B model. This suggests that the large model with more pre-training data has an advantage over smaller one when trained with less noisy labels. Further details of the WER of the pseudo-labels at each training stage is discussed in the following section.


\begin{table}[t]
    \footnotesize
    \centering
    \caption{Performance comparison of wav2vec-large (w2v-L) and wav2vec-base (w2v-B) models at each stage.}
    \label{tab:w2vcomp}
    \begin{tabular}{l|c c|c c|c c}
        \hline
        \hline
        & \multicolumn{2}{c|}{$S1$} & \multicolumn{2}{c|}{$S2$} & \multicolumn{2}{c}{$S3$}\\
        \hline
        \multirow{2}{*}{\bf Test sets} & \bf w2v-B & \bf w2v-L & \bf w2v-B & \bf w2v-L & \bf w2v-B & \bf w2v-L \\
        \cline{2-7}
        & \multicolumn{6}{c}{\textit{w/o LM}}\\
        \hline
        eval00 & \bf 31.5 & 32.0 & 24.7 & \bf 24.3 & 22.9 & \bf 21.0 \\
        CH & 36.4 & \bf 36.0 & 29.2 & \bf 28.1 & 27.6 & \bf 24.5\\
        SB & \bf 26.3 & 27.8 & \bf 20.1 & 20.3 & 17.9 & \bf 17.4\\
        \hline
        & \multicolumn{6}{c}{\textit{w/ LM (OOD)}}\\
        \hline
        eval00 & \bf 26.0 & 26.2 & 22.4 & \bf 21.3 & 20.5 & \bf 19.6\\
        CH & 30.6 & \bf 30.2 & 26.7 & \bf 25.0 & 24.8 & \bf 23.1\\
        SB & \bf 21.2 & 22.0 & 17.9 & \bf 17.4 & 16.0 & \bf 16.0\\
        \hline
        \hline
    \end{tabular}
    \vspace{-3mm}
\end{table}

\begin{table}[t]
    \footnotesize
    \centering
    \caption{WER of the teacher and student models on SWBD training data computed during inference. For the 'Teachers' model, WER is computed after selection strategy.}
    \label{tab:TrainError}
    \begin{tabular}{l|c c}
        \hline
        \hline
        \multirow{2}{*}{\bf Model} & \multicolumn{2}{c}{\bf WER of Training data} \\
        & w/o LM & w/ LM (OOD) \\
        \hline
        Teachers & 41.9 & 38.8 \\ 
        $S1$ & 32.4 & 26.8\\
        $S2$ & 25.5 & 22.9\\
        $S3$ & 22.9 & 22.0\\
        \hline
        \hline
    \end{tabular}
\vspace{-3mm}
\end{table}

\vspace{-2mm}
\subsection{Analysis of selection strategy and pseudo-labels}
\label{sec:analysis}
This section aims to provide insights into the the performance of selection strategy and WER of pseudo-labels at teachers and student stages. These analysis are done for the SWBD training data during inference. With the help of oracle selection it is computed that almost 75.5\% utterances are correctly selected by selection method which lead to the 32\% WER for $S1$ compared to 41.8\% of LS360. Student model trained with oracle selection (100\% correct) $S_{Or}$ is able to achieve WER of 27.4\% as shown in table \ref{tab:multi-stage}. This indicates that improvement in the selection method would eventually lead to better first stage student and it would be further propagated to the subsequent students.

Table \ref{tab:TrainError} shows the WER of teacher and student models on SWBD training data. Results in this table may help to understand how much pseudo-labels are improving at each training stage leading to better subsequent student model. The results in this table are presented with and without using beam-search and OOD LM. As discussed earlier, pseudo-labels only acquired through beam-search and OOD LM are used in training the next stage model. 
The results w/o LM are helpful to understand the contribution of OOD LM in improving the pseudo-labels. 
In table \ref{tab:TrainError}, WER of `Teachers' is computed after the selection strategy. The WER of teacher models after selection is 41.93\% which is improved to 38.8\% using OOD LM. This means that the $S1$ is trained with pseudo-labels having WER of 38.8\%. The WER of the $S1$ model on the SWBD training set is 32.4\% (w/o LM) and 26.8\% (w/ OOD LM). It can be seen that pseudo-labels provided by $S1$ are significantly improved compared to the teachers i.e. 41.9\% (w/o LM) to 32.4\% (w/o LM) with 9.5\% absolute gain. Using OOD LM the difference between `Teachers' and $S1$ is increased to 12\% with WER of 26.8\% (w/ OOD LM). At this stage, there is a significant improvement in the error of pseudo-labels. Similarly, at the $S2$ and the $S3$ models the WER become 22.9 and 22.0 with OOD LM. Finally, it can be seen that at the $S3$ model there is no significant improvement in pseudo-labels. It means no further improvements is expected if more stages are trained afterwards. 

\vspace{-4mm}
\section{Conclusion}
\vspace{-2mm}
This paper presented multi-stage student training method in combination with ensemble teacher models for improved unsupervised domain adaptation. The student models evaluated on SWBD test sets have shown significant improvement starting from the first stage to the third stage student. Each student was trained using the pseudo-labels generated by the previous teacher models. The progressive improvement at each stage in terms of WER has shown the effectiveness of employing multi-stage training method. Furthermore, it was evaluated that the pseudo-labels with higher WER do not show any advantage of using wav2vec-large model compared to the wav2vec-base. However, as the labels get better in subsequent stages the model based on wav2vec-large gives better performance. 

\vspace{-2mm}
\section{Acknowledgment}
\vspace{-3mm}
This work was conducted at the LivePerson centre of
Speech and Language Technology at the university of Sheffield
which is supported by LivePerson, Inc.

\ninept
\bibliographystyle{ieeetr}
\bibliography{refs}

\end{document}